\documentclass[conference]{IEEEtran}

\usepackage{multicol}
\usepackage{etoolbox}
\makeatletter
\patchcmd{\@makecaption}
  {\scshape}
  {}
  {}
  {}
\makeatletter
\patchcmd{\@makecaption}
  {\\}
  {.\ }
  {}
  {}
\makeatother

\usepackage{amsfonts}
\usepackage{mathrsfs}
\usepackage{mathtools}
\usepackage{amsfonts}
\usepackage{amssymb}
\usepackage{graphicx}
\usepackage{epsfig}
\usepackage{psfrag}
\usepackage{amsmath}
\usepackage{array}
\usepackage{cases}
\usepackage{eufrak}
\usepackage{cite,graphicx,amsmath,amssymb,color}
\usepackage{algorithmic}
\usepackage{algorithm}
\usepackage{subfigure}
\usepackage{bm}
\usepackage{multirow}
\usepackage{threeparttable}
\usepackage{array}
\usepackage{makecell}

\DeclareMathOperator*{\argmax}{argmax}

\newtheorem{Cla}{Claim}
\newtheorem{Thm}{Theorem}

\newtheorem{Prob}{Problem}

\IEEEoverridecommandlockouts
\begin{document}
\title{\textcolor{black}{Optimal} Transmission of Multi-Quality Tiled 360 VR Video in MIMO-OFDMA Systems}
\author{\IEEEauthorblockN{Chengjun Guo,  Ying Cui,  Zhi Liu,  and Derrick Wing Kwan Ng}\thanks{C. Guo and Y. Cui are with Shanghai Jiao Tong University, China. Z. Liu is with the University of Electro-Communications, Japan.  D. W. K. Ng is with the University of New South Wales, Australia. (Corresponding author: Ying Cui.)}}

\maketitle


\begin{abstract}
In this paper, we study \textcolor{black}{the} optimal  transmission of a multi-quality tiled 360 virtual reality (VR) video
from a multi-antenna server (e.g., access point or base \allowbreak station)~to~multiple single-antenna users in a multiple-input multiple-output (MIMO)-orthogonal frequency division multiple access \allowbreak(OFDMA) system.
\textcolor{black}{ We minimize the total transmission power
 with respect to the subcarrier allocation~constraints, rate allocation constraints, and successful transmission constraints,
 by optimizing the beamforming vector and subcarrier, transmission power and rate allocation.}
The formulated resource allocation problem is a challenging mixed
discrete-continuous optimization problem.
We obtain  an asymptotically optimal solution in the case of a large antenna array, and a suboptimal solution in the general case.
\textcolor{black}{As far as we know}, this is the first work providing optimization-based design for 360 VR video transmission in MIMO-OFDMA systems.
Finally, \textcolor{black}{by numerical results, we show that the proposed solutions  achieve significant  improvement in performance compared to the existing solutions.}
\end{abstract}

\section{Introduction}
\textcolor{black}{By using an omnidirectional camera to capture a scene of interest in all directions at the same time, a 360 virtual reality (VR) can be generated.}
In many VR applications, e.g., VR gaming, VR concert, and VR military training,
a 360 VR video has to be transmitted to multiple users simultaneously.
Transmitting a 360 VR video over wireless networks enables users to experience immersive environments  without
geographical or behavioral restrictions.
\textcolor{black}{In this paper, the main focus is on
optimally transmitting
 a  360 VR video
 in a multi-user wireless network.}

A 360 VR video has a much larger size  than  a traditional video.
\textcolor{black}{When watching a 360 VR video, a user is  perceiving it} from only one viewing direction  at any time, which corresponds to one part of the 360 VR video, known as field-of-view (FoV).
\textcolor{black}{Tiling technique is widely used to improve the transmission efficiency for 360 VR videos \cite{deterfov}.}  Transmitting the set of tiles \textcolor{black}{which cover  predicted FoVs} can reduce the required communication resources, without \textcolor{black}{affecting} the  quality of experience. \textcolor{black}{When transmitting a tiled 360 VR video   in a multi-user wireless network, if there exists a tile required by
multiple
users concurrently,
multicast opportunities can be utilized to improve transmission efficiency.}
Recently, \cite{guo_tdma,guo_ofdma,multi-user1,multi-user3,long,long2}
study \textcolor{black}{streaming} of a tiled 360 VR video
from a single-antenna server to multiple single-antenna users \textcolor{black}{in wireless networks},
where multicast
opportunities are exploited.
In particular, in
our previous works \cite{guo_tdma,guo_ofdma},  \textcolor{black}{the optimal transmission
of a single-quality tiled 360 VR video is studied}.
\textcolor{black}{In practice,} pre-encoding each tile into multiple representations with different quality levels allows quality
\textcolor{black}{adaptation} according to users' channel conditions.
In \cite{multi-user1,multi-user3,long,long2},
\textcolor{black}{the optimal  transmission of a multi-quality tiled 360 VR video is considered, with the main focus on the optimization of
quality level selection for each tile.}

Despite the fruitful research in the literature, the performance of wireless transmission of a tiled 360 VR video is still limited. In fact, the results in
\cite{guo_tdma,guo_ofdma,multi-user1,multi-user3,long,long2} all consider single-antenna
servers, which cannot exploit  \textcolor{black}{the} spatial degrees of freedom.
\textcolor{black}{The performance of wireless systems can be significantly improved by deploying multiple antennas at a server  and designing efficient beamformers.}
 \textcolor{black}{Among various multi-antenna technologies,} MIMO-OFDMA is the dominant air interface for  5G broadband wireless communications,
as it can provide more reliable communications at high speeds.
 \textcolor{black}{For instance,} in
\cite{mo2,mo3}, the authors consider
 multi-group multicast in MIMO-OFDMA systems.
Specifically,
In \cite{mo2}, the subcarrier and power allocation is considered to maximize the  \textcolor{black}{system} sum rate.
However, the solution proposed in \cite{mo2} is heuristic, and hence has no performance guarantee.
In \cite{mo3}, the authors study the optimization of beamformers to minimize the total transmission power, and obtain a
stationary point of the beamforming design problem  based on successive convex approximation. Note that in \cite{mo3}, messages on each subcarrier have different beamformers,   resulting
in a substantial increase in the number of variables and hence the computational complexity for
solving the optimization problem.

In this paper, we consider \textcolor{black}{the} optimal  transmission of a multi-quality tiled 360 VR video \textcolor{black}{in \textcolor{black}{a} MIMO-OFDMA system.} With more advanced physical layer techniques \textcolor{black}{than those in \cite{guo_tdma,guo_ofdma,multi-user1,multi-user3,long,long2}}, we expect \textcolor{black}{the} stringent requirements for 360 VR video transmission to be better satisfied.
\textcolor{black}{We  minimize the total transmission power by optimizing the beamforming vectors and subcarrier, transmission power, and rate allocation,}  under the subcarrier allocation constraints, rate allocation constraints, and successful transmission constraints.
\textcolor{black}{This problem is a  mixed discrete-continuous optimization problem, and is very challenging.}
We obtain its asymptotically optimal solution \textcolor{black}{in the special case of a large antenna array}, by exploiting decomposition, continuous relaxation, and \textcolor{black}{Karush-Kuhn-Tucker (KKT)} conditions. We also obtain a suboptimal solution in the general case,  \textcolor{black}{applying} continuous relaxation and \textcolor{black}{difference of convex (DC)} programming.
Note that \textcolor{black}{the transmission of a tiled 360 VR video in this paper} can be viewed as
multi-group multicast.
Previous works studying multi-group multicast in MIMO-OFDMA systems \textcolor{black}{do} not \textcolor{black}{investigate} \textcolor{black}{the special case of a large antenna array where an asymptotically optimal solution} can be obtained \cite{mo2,mo3}.
Note that the proposed formulation in this paper, with one beamforming vector for each subcarrier, can achieve the same performance as the formulation in \cite{mo3} but with much lower computational complexity in the general case.
Finally, numerical results show substantial gains \textcolor{black}{achieved} by the proposed
solutions over existing schemes.


\section{System Model}
\begin{figure}[t]
\begin{center}
\subfigure[\small{Multi-quality tiled 360 VR video required by multiple users. }] 
 {\resizebox{6cm}{!}{\includegraphics{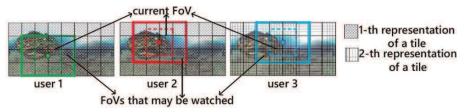}}}
 \subfigure[\small{Transmission of a multi-quality tiled 360 VR video.}]
 {\resizebox{6cm}{!}{\includegraphics{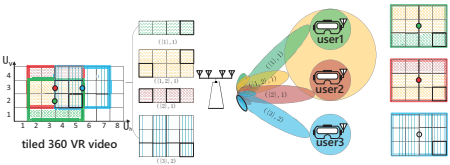}}}
\vspace*{-0.07cm}
 \end{center}
  \vspace*{-0.2cm}
     \caption{\small{System model of transmission of a multi-quality tiled 360 VR video. $K=3$, $\mathbf{r}=(1,1,2),$  $U_h\times U_v=8\times4$, $M=4$,
     $\mathcal{G}_1=\{(2,1),(3,1),(4,1),(5,1),(2,2),(3,2),(4,2),(5,2),(2,3),(3,3)$,
     $(4,3),(5,3)\}$,
$\mathcal{G}_2=\{(2,2),(3,2),(4,2),(5,2),(2,3),(3,3),(4,3)$,
$(5,3),(2,4),(3,4),(4,4),(5,4)\}$, $\mathcal{G}_3=\{(4,2),(5,2),(6,2),(7,2)$,
$(4,3),(5,3),(6,3),(7,3),(4,4),(5,4),(6,4),(7,4)\}$,
 $\mathcal{I}=\{\{1\},\{2\},\{3\},\{1,3\},\{2,3\},\{1,2,3\}\}$,
 $\mathcal{P}_{\{1\}}=\{(2,1),(3,1),(4,1),(5,1)\}$,
 $\mathcal{P}_{\{2\}}=\{(2,4),(3,4)\}$,
 $\mathcal{P}_{\{3\}}=\{(6,2),(6,3),(6,4),(7,2),(7,3),(7,4)\}$,
 $\mathcal{P}_{\{1,2\}}=\{(2,2),(2,3),(3,2),(3,3)\}$,
 $\mathcal{P}_{\{2,3\}}=\{(4,4),(5,4)\}$,
 $\mathcal{P}_{\{1,2,3\}}=\{(4,2),(4,3),(5,2),(5,3)\}$,
 \textcolor{black}{$\mathcal{K}_{\{1\},1}=\{1\},$
 $\mathcal{K}_{\{1,2\},1}=\{1,2\},$
 $\mathcal{K}_{\{2\},1}=\{2\},$
 $\mathcal{K}_{\{3\},2}=\{3\}.$}
 }}
\label{fig:system-model}
\vspace*{-0.5cm}
\end{figure}

We consider the streaming of a multi-quality tiled 360 VR video from a server (e.g., access point or base station) to $K$ users in an MIMO-OFDMA system as illustrated in Fig.~\ref{fig:system-model}.\footnote{We adopt a  multi-quality tiled 360 VR video model which is similar to those  in  our previous works \cite{guo_tdma,guo_ofdma,long,long2}, and the details are presented here for completeness.} The server is equipped with $M$ transmit antennas and each user wears a single-antenna VR headset. Denote $\mathcal{K}\triangleq\{1,\ldots,K\}$ as the set of user indices.
When a VR user is interested in one viewing direction of a 360 VR video, \textcolor{black}{the user watches} a rectangular FoV of size $F_h\times F_v$ (in rad$\times$rad), the center of which is referred to as the viewing direction.
\textcolor{black}{In addition, a user can freely switch views when watching a 360 VR video.}

\textcolor{black}{Tiling is adopted to improve transmission efficiency of the 360 VR video.} In particular, \textcolor{black}{the 360 VR video is divided into} multiple rectangular segments of the same size, which are referred to as tiles. Let $U_h$ and $U_v$ denote the numbers of segments in each \textcolor{black}{row and column}, respectively. Define $\mathcal{U}_h\triangleq\{1,\ldots,U_h\}$ and $\mathcal{U}_v\triangleq\{1,\ldots,U_v\}$.
The $(u_h,u_v)$-th tile is referred to as the tile in the $u_h$-th row and the $u_v$-th column, for all $u_h\in\mathcal{U}_h$ and $u_v\in\mathcal{U}_v$.
Considering user heterogeneity (\textcolor{black}{e.g., display resolutions of devices, channel conditions, etc.}),  \textcolor{black}{each tile  is pre-encoded} into $L$ representations corresponding to $L$ quality levels using
\textcolor{black}{High Efficiency Video Coding (HEVC)}, as in  Dynamic Adaptive Streaming over HTTP (DASH).
\textcolor{black}{Denote $\mathcal{L}\triangleq\{1,\ldots,L\}$ as the set of quality levels.}
For all $l\in\mathcal{L}$, the $l$-th representation of each tile corresponds to the $l$-th lowest quality. For ease of exposition, \textcolor{black}{we assume that the encoding rates of the tiles with the same quality level are identical}. Let $D_l$ (in bits/s) denote the encoding rate of the $l$-th representation of a tile.
Note that $D_1<D_2<\ldots<D_L$.
\textcolor{black}{We consider} the duration of the playback time of \textcolor{black}{one group of pictures (GOP)},\footnote{\textcolor{black}{The playback time of one GOP is $0.06$-$1$ seconds in general.}} over which the FoV of each user does not change. Let \textcolor{black}{$r_k\in\mathcal{L}$} denote the quality level for the FoV of user $k\in\mathcal{K}$.
\textcolor{black}{Due} to the video coding structure, $\mathbf{r}\triangleq(r_k)_{k\in\mathcal{K}}$ should not change during the considered time duration.


As in \cite{guo_tdma,guo_ofdma,long}, suppose the FoV of each user has been predicted and we focus on \textcolor{black}{the} transmission of \textcolor{black}{a} multi-quality tiled 360 VR video. Let $\mathcal{G}_k$ denote the set of indices of the tiles \textcolor{black}{which} need to be transmitted to user $k$ and let $\mathcal{G}\triangleq\bigcup_{k\in\mathcal{K}}\mathcal{G}_k$ denote the set of indices of the tiles \textcolor{black}{which} need to be transmitted considering all $K$ users.\footnote{The proposed framework does not depend on any particular method
for determining the set of tiles \textcolor{black}{which are} to be transmitted to each user \cite{deterfov,guo_tdma,guo_ofdma}.}
 For all $\mathcal{S}\subseteq\mathcal{K}$, $\mathcal{S}\neq\emptyset$,
 let $\mathcal{P}_{\mathcal{S}}\triangleq\left(\bigcap_{k\in\mathcal{S}}\mathcal{G}_k
 \right)\bigcap\left(\mathcal{G}-\bigcup_{k\in\mathcal{K}\setminus\mathcal{S}}\mathcal{G}_k\right)$
denote the set of indices of the tiles that are needed by all users in $\mathcal{S}$ and are not needed by the users in $\mathcal{K}\setminus\mathcal{S}$.
Then $\mathcal{P}\triangleq\{\mathcal{P}_{\mathcal{S}}|\mathcal{P}_{\mathcal{S}}
\neq\emptyset,~\mathcal{S}\subseteq\mathcal{K},~\mathcal{S}\neq\emptyset\}$ forms
a partition of $\mathcal{G}$
and $\mathcal{I}\triangleq\{\mathcal{S}|\mathcal{P}_{\mathcal{S}}\neq\emptyset,
~\mathcal{S}\subseteq\mathcal{K},~\mathcal{S}\neq\emptyset\}$
specifies the user sets corresponding to the partition.
Let $\mathcal{I}_k\triangleq\{\mathcal{S}|\mathcal{S}\subseteq\mathcal{I},k\in\mathcal{S}\}$,  $k\in\mathcal{K}$.
The tiles in $\mathcal{P}_{\mathcal{S}}$, $\mathcal{S}\in\mathcal{I}_k$ are \textcolor{black}{required} by user $k$.
We jointly consider the tiles in each set, \textcolor{black}{rather than treat
them separately, to
reduce the complexity for transmission and resource allocation}.
\textcolor{black}{For all $\mathcal{S}\in\mathcal{I}$, let $\mathcal{L}_{\mathcal{S}}\triangleq\{r_k|k\in\mathcal{S}\}$.}
For all \textcolor{black}{$l\in\mathcal{L}_{\mathcal{S}}$} and $\mathcal{S}\subseteq\mathcal{K}$,
\textcolor{black}{the encoding (source coding) bits of}
the $l$-th representations of the tiles in $\mathcal{P}_{\mathcal{S}}$ \textcolor{black}{are ``aggregated'' into}
one message indexed by \textcolor{black}{$(\mathcal{S},l)$}, which is transmitted at most once to the users in $\mathcal{S}$ that will utilize it, to improve transmission efficiency.
\textcolor{black}{For all $\mathcal{S}\in\mathcal{I}$ and $l\in\mathcal{L}_{\mathcal{S}}$,
let $\mathcal{K}_{\mathcal{S},l}\triangleq\{k\in\mathcal{S}|r_k=l\}$.}
If there is only one user \textcolor{black}{in $\mathcal{K}_{\mathcal{S},l}$}, the transmission of message \textcolor{black}{$(\mathcal{S},l)$} corresponds to unicast;
\textcolor{black}{and if} there are multiple users \textcolor{black}{in $\mathcal{K}_{\mathcal{S},l}$}, the transmission of message \textcolor{black}{$(\mathcal{S},l)$} corresponds to multicast.
Thus, the transmission of the \textcolor{black}{multi-quality tiled}  360 VR video to the $K$ users may involve
\textcolor{black}{both} unicast and multicast.
An illustration example can be seen in Fig.~\ref{fig:system-model} (b). \textcolor{black}{In this example, the server multicasts message $(\{1,2\},1)$ to user 1 and user 2.}

\textcolor{black}{
Let $\mathcal{N}\triangleq\{1,\ldots,N\}$, where $N$ is the number of subcarriers.}
The bandwidth of each subcarrier is  $B$ (in Hz).
\textcolor{black}{We} assume block fading, i.e., \textcolor{black}{the small-scale channel fading
coefficients do} not change within one frame.
Let $\mathbf{h}_{n,k}\in\mathbb{C}^{M\times1}$ denote the
small scale fading coefficient between the server and user $k$ on subcarrier $n$.
Denote $\mathbf{h}\triangleq (\mathbf{h}_{n,k})_{n\in\mathcal{N},k\in\mathcal K} $ as the  system channel state.
\textcolor{black}{Assume} that the server is aware of $\mathbf{h}$, by channel estimation.
Let $\beta_k>0$ denote the \textcolor{black}{large-scale channel fading gain} between the server and user $k$, which remains constant during the considered time duration  and is known to the server.

\textcolor{black}{Denote $\mu_{\mathcal{S},l,n}\in\{0,1\}$  as} the subcarrier assignment indicator
for subcarrier $n$ and message \textcolor{black}{$(\mathcal{S},l)$},
where $\mu_{\mathcal{S},l,n}=1$ indicates that subcarrier $n$ is assigned to transmit \textcolor{black}{the symbols} for message \textcolor{black}{$(\mathcal{S},l)$}, and $\mu_{\mathcal{S},l,n}=0$ otherwise.
\textcolor{black}{For  ease of implementation, we assume that} each subcarrier is assigned to transmit \textcolor{black}{symbols} \textcolor{black}{of only} one message. \textcolor{black}{Note that it is a commonly adopted assumption \cite{mo2}.}
 Thus, subcarrier allocation constraints is given by
\begin{equation}
    \mu_{\mathcal{S},l,n} \in\{0,1\},~\mathcal{S}\in\mathcal{I},l\in\mathcal{L}_{\mathcal{S}},n\in\mathcal{N},\label{cst:mu1}
\end{equation}
\begin{equation}
    \sum\nolimits_{\mathcal{S}\in\mathcal{I}}\sum\nolimits_{l\in\mathcal{L}_{\mathcal{S}}}\mu_{\mathcal{S},l,n} =1,~n\in\mathcal{N}.\label{cst:mu2}
\end{equation}

\textcolor{black}{To capture the scaling of the transmission power with $M$ for studying the optimal power allocation at large $M$,} let $\frac{\eta_{\mathcal{S},l,n}}{M}$ denote the transmission power for  \textcolor{black}{the symbols} for message \textcolor{black}{$(\mathcal{S},l)$} on  subcarrier $n$, where
\begin{equation}
 \eta_{\mathcal{S},l,n} \geq0,~\mathcal{S}\in\mathcal{I},l\in\mathcal{L}_{\mathcal{S}},n\in\mathcal{N}.\label{cst:eta}
\end{equation}
The total transmission power is $\sum\limits_{n\in\mathcal{N}}\sum\limits_{\mathcal{S}\in\mathcal{I}}\sum\limits_{l\in\mathcal{L}_{\mathcal{S}}}\frac{\mu_{\mathcal{S},l,n} \eta_{\mathcal{S},l,n}}{M}.$

Suppose \textcolor{black}{the} subcarrier $n$ is assigned to transmit \textcolor{black}{the symbols} for message \textcolor{black}{$(\mathcal{S},l)$}. Let $s_{\mathcal{S},l,n}$ represent \textcolor{black}{the symbols} for message \textcolor{black}{$(\mathcal{S},l)$} transmitted on subcarrier $n$.
Assume $\mathbb{E}[|s_{\mathcal{S},l,n}|^2]=1$.
Let $\mathbf{w}_{n}\in\mathbb{C}^{M\times1}$ denote the  beamforming vector for the message transmitted on subcarrier $n$,
where
\begin{equation}
\|\mathbf{w}_{n} \|=1,~n\in\mathcal{N}.\label{cst:w}
\end{equation}
The received signal at user $k$ on subcarrier $n$ is given by
\begin{align}
&y_{\mathcal{S},l,k,n}=\sqrt{\frac{\beta_{k}\eta_{\mathcal{S},l,n}}{M}}\mathbf{h}_{n,k} ^{H}\mathbf{w}_{n}s_{\mathcal{S},l,n}+z_{n,k},~k\in\mathcal{K},n\in\mathcal{N},\nonumber
\end{align}
where $z_{n,k}\sim\mathcal{CN}(0,\sigma^2)$ represents the noise at user $k$ on subcarrier $n$.
\textcolor{black}{Capacity achieving code is adopted to obtain design insights} \cite{luo2006}. The
maximum transmission rate for \textcolor{black}{the symbols} for message \textcolor{black}{$(\mathcal{S},l)$} to user $k\in\mathcal{S}$ on subcarrier $n$  is given by $B\log_2\left(1+
    \frac{\beta_{k}\eta_{\mathcal{S},l,n}|\mathbf{h}_{n,k}^H\mathbf{w}_{n}|^2}
    {M\sigma^2}\right)$ (\textcolor{black}{in bit/s}).

Let $c_{\mathcal{S},l,n}$   denote the transmission rate for  \textcolor{black}{the symbols} for message \textcolor{black}{$(\mathcal{S},l)$}  on  subcarrier $n$,  where
\begin{equation}
c_{\mathcal{S},l,n}\geq0,~\mathcal{S}\in\mathcal{I},l\in\mathcal{L}_{\mathcal{S}},n\in\mathcal{N}.\label{cst:c}
\end{equation}
To guarantee that message $(\mathcal{S},l)$
can be successfully transmitted to each user $k\in\mathcal{K}_{\mathcal{S},l}$
on subcarrier $n$,  \textcolor{black}{we have}
\begin{align}
     &\mu_{\mathcal{S},l,n}B\log_2\left(1+
    \frac{\beta_{k}\eta_{\mathcal{S},l,n}|\mathbf{h}_{n,k}^H\mathbf{w}_{n}|^2}
    {M\sigma^2}\right)\geq c_{\mathcal{S},l,n} ,\nonumber\\
&   ~~~~~~~~~~~~~~~~~~~~~~~~~~~ \mathcal{S}\in\mathcal{I},l\in\mathcal{L}_{\mathcal{S}},k\in\mathcal{K}_{\mathcal{S},l},n\in\mathcal{N}.\label{cst:minrate}
\end{align}
\textcolor{black}{To avoid stalls} during the video playback for message $(\mathcal{S},l)$,
\textcolor{black}{the  transmission rate constraint is given by}
\begin{equation}
\sum\limits_{n\in\mathcal{N}}c_{\mathcal{S},l,n}\geq |\mathcal{P}_{\mathcal{S}}|D_{l},~\mathcal{S}\in\mathcal{I},l\in
    \mathcal{L}_{\mathcal{S}},\label{cst:averate}
\end{equation}
where $|\mathcal{P}_{\mathcal{S}}|$ denotes the number of tiles in $\mathcal{P}_{\mathcal{S}}$.




\section{Total Transmission Power Minimization}\label{section:minp}

For convenience, denote
$\bm{\mu} \triangleq(\mu_{\mathcal{S},l,n} )_{\mathcal{S}\in\mathcal{I},l\in\mathcal{L}_{\mathcal{S}},n\in\mathcal{N}}$,
$\bm{\eta} \triangleq(\eta_{\mathcal{S},l,n} )_{\mathcal{S}\in\mathcal{I},l\in\mathcal{L}_{\mathcal{S}},n\in\mathcal{N}}$,
 $\mathbf{c} \triangleq(c_{\mathcal{S},l,n} )_{\mathcal{S}\in\mathcal{I},l\in\mathcal{L}_{\mathcal{S}},n\in\mathcal{N}}$,
 and $\mathbf{w} \triangleq(\mathbf{w}_{n} )_{n\in\mathcal{N}}$.
Given $(\mathcal{G}_k)_{k\in\mathcal{K}}$ and  $\mathbf{r}$, we would
like to  \textcolor{black}{minimize the total transmission power subject to the constraints in
\eqref{cst:mu1}-\eqref{cst:averate}, by optimizing the normalized beamforming vectors $\mathbf{w} $ and the subcarrier $\bm{\mu} $, power $\bm{\eta} $ and rate $\mathbf{c} $ allocation.}
\begin{Prob}[Total Transmission Power Minimization]\label{P1}
\begin{align}
E^{\star}\triangleq\min_{\bm{\mu},\bm{\eta},\mathbf{c},\mathbf{w}}~
&\frac{1}{M}\sum\nolimits_{n\in\mathcal{N}}\sum\nolimits_{\mathcal{S}\in\mathcal{I}}\sum\nolimits_{l\in\mathcal{L}_{\mathcal{S}}}\mu_{\mathcal{S},l,n} \eta_{\mathcal{S},l,n}\nonumber\\
    \mathrm{s.t.} ~~
    &\eqref{cst:mu1},~\eqref{cst:mu2},~\eqref{cst:eta},~\eqref{cst:w},~\eqref{cst:c},~\eqref{cst:minrate},~\eqref{cst:averate}.\nonumber
\end{align}
\end{Prob}

Problem~\ref{P1} is a challenging discrete-continuous optimization problem.
\textcolor{black}{In Section~\ref{sec:p1-sp} and Section~\ref{sec:p1-dc}, we solve Problem~\ref{P1} in a special case} and the general case, respectively.

\subsection{Asymptotically Optimal Solution}\label{sec:p1-sp}
In this subsection, we
solve Problem~\ref{P1} in \textcolor{black}{the special case where} the server is equipped with  a
large antenna array, by solving the following equivalent problem of Problem~\ref{P1}.

%

\begin{Prob}[Equivalent Problem of Problem~\ref{P1}]\label{EP1}
\begin{align}
&\min_{\bm{\mu},\mathbf{P} }~\frac{1}{M}\sum\nolimits_{n\in\mathcal{N}}\sum\nolimits_{\mathcal{S}\in\mathcal{I}}\sum\nolimits_{l\in\mathcal{L}_{\mathcal{S}}}P_{\mathcal{S},l,n} \nonumber\\
    &\mathrm{s.t.} ~~
    \eqref{cst:mu1},~\eqref{cst:mu2},\nonumber\\
    &\textcolor{black}{P_{\mathcal{S},l,n}\geq0,~\mathcal{S}\in\mathcal{I},l\in\mathcal{L}_{\mathcal{S}},n\in\mathcal{N},}\label{cst:p}\\
&\sum\nolimits_{n\in\mathcal{N}}\mu_{\mathcal{S},l,n}B\log_2\left(1+
    \frac{P_{\mathcal{S},l,n}}
    {\mu_{\mathcal{S},l,n}Q_{\mathcal{S},l,n}^{\dagger}}\right)\geq |\mathcal{P}_{\mathcal{S}}|D_{l},\nonumber\\
    &~~~~~~~~~~~~~~~~~~~~~~~~~~~~~~~~~~\mathcal{S}\in\mathcal{I},l\in\mathcal{L}_{\mathcal{S}},k\in\mathcal{K}_{\mathcal{S},l},\label{cst:sumraten}
\end{align}
\textcolor{black}{where $\mathbf{P}\triangleq(P_{\mathcal{S},l,n})_{\mathcal{S}\in\mathcal{I},l\in\mathcal{L}_{\mathcal{S}},n\in\mathcal{N}}$}
and $Q_{\mathcal{S},l,n}^{\dagger}$ is given by the \textcolor{black}{optimal value of the} following problem.
\textcolor{black}{Denote $(\bm{\mu}^{\dagger} ,\mathbf{P}^{\dagger} )$ as} an optimal solution of Problem~\ref{EP1}.
\end{Prob}
\begin{Prob}[\textcolor{black}{Subproblem of Problem~\ref{EP1}}]\label{BF1}
For all $\mathcal{S}\in\mathcal{I}$,~$l\in\mathcal{L}_{\mathcal{S}}$ and  $n\in\mathcal{N}$,
\begin{align}
Q_{\mathcal{S},l,n}^{\dagger}\triangleq&\min_{\mathbf{V}_{\mathcal{S},l,n}\in\mathbb{C}^{M\times M}}\text{trace}(\mathbf{V}_{\mathcal{S},l,n})
    \nonumber\\
    \mathrm{s.t.} ~~&\frac{\text{trace}(\beta_k\mathbf{h}_{k,n} \mathbf{h}_{k,n} ^H\mathbf{V}_{\mathcal{S},l,n})}{M\sigma^2}\geq1,~k\in\mathcal{K}_{\mathcal{S},l},\nonumber\\
    &\mathbf{V}_{\mathcal{S},l,n}\succeq\mathbf{0},\nonumber\\
    &\text{rank}(\mathbf{V}_{\mathcal{S},l,n})=1.\label{cst:rank1}
\end{align}
\textcolor{black}{Denote $\mathbf{V}_{\mathcal{S},l,n}^{\dagger} $ as} an optimal solution of Problem~\ref{BF1},
which can be written as $\mathbf{V}_{\mathcal{S},l,n}^{\dagger} =\mathbf{v}^{\dagger}_{\mathcal{S},l,n}
(\mathbf{v}^{\dagger}_{\mathcal{S},l,n} )^H$ for some $\mathbf{v}^{\dagger}_{\mathcal{S},l,n} \in\mathbb{C}^{M\times1}$.
\end{Prob}

\textcolor{black}{By making use of} structures of Problem~\ref{P1}, Problem~\ref{EP1}, and Problem~\ref{BF1}, we have the following result.\footnote{Please refer to \cite{report} for the proof.}
\begin{Thm}[Equivalence between Problem~\ref{P1} $\&$ Problem~\ref{EP1}]\label{lem:eqp1}
\textcolor{black}{Problem~\ref{P1} and Problem~\ref{EP1} have the same optimal value.}
In addition, $(\bm{\mu}^{\dagger},\bm{\eta}^{\dagger},\mathbf{c}^{\dagger},
\mathbf{w}^{\dagger})$
is an optimal solution of Problem~\ref{P1}, where
$\bm{\eta}^{\dagger}=\mathbf{P}^{\dagger}$, $\mathbf{w}^{\dagger}_{n} =\sum\nolimits_{\mathcal{S}\in\mathcal{I}}\sum\nolimits_{l\in\mathcal{L}_{\mathcal{S}}}\mu_{\mathcal{S},l,n}^{\dagger}\frac{\mathbf{v}^{\dagger}_{\mathcal{S},l,n} }{\sqrt{Q_{\mathcal{S},l,n}^{\dagger}}}$, $n\in\mathcal{N}$,
and $\mathbf{c}^{\dagger}\triangleq(\mathbf{c}^{\dagger}_{\mathcal{S},l,n})_{\mathbf{S}\in\mathcal{I},l\in\mathcal{L}_{\mathcal{S}},n\in\mathcal{N}}$
with $c_{\mathcal{S},l,n}^{\dagger} =\mu_{\mathcal{S},l,n}^{\dagger}B\log_2\left(1+
    \frac{P_{\mathcal{S},l,n}^{\dagger} }
    {Q_{\mathcal{S},l,n}^{\dagger} }\right)$, $\mathcal{S}\in\mathcal{I},l\in\mathcal{L}_{\mathcal{S}},n\in\mathcal{N}.$
\end{Thm}

\textcolor{black}{According to Theorem~\ref{lem:eqp1}, to obtain an optimal solution of Problem~\ref{P1},
we can first obtain $\mathbf{w}^{\dagger}$ by solving Problem~\ref{BF1} and then obtain $\bm{\mu}^{\dagger}$, $\bm{\eta}^{\dagger}$, and $\mathbf{c}^{\dagger}$ by solving
Problem~\ref{EP1}.}
\textcolor{black}{Notice that}
 Problem~\ref{BF1} is a nonconvex problem
due to the rank-one constraint in  \eqref{cst:rank1}, \textcolor{black}{while}
Problem~\ref{EP1} is a nonconvex problem \textcolor{black}{because of} the binary constraints in
\eqref{cst:mu1}.
\textcolor{black}{Both problems are quite challenging.}

To obtain an asymptotically optimal solution of Problem~\ref{BF1} for  large $M$,
we explicitly write the optimal value of Problem~\ref{BF1}
as a function of $M$, i.e., $Q_{\mathcal{S},l,n}^{\dagger(M)} $.
Following \textcolor{black}{a similar approach for the proofs for} \textcolor{black}{Theorem~1 and Theorem~3}  in \cite{xiang}, we \textcolor{black}{have} the following result.

\begin{Thm}[Asymptotically Opt. Solution of Problem~\ref{BF1}]\label{thm:M}
For all $\mathcal{S}\in\mathcal{I}$, $l\in\mathcal{L}_{\mathcal{S}}$, and
$n\in\mathcal{N}$, $\mathbf{V}_{\mathcal{S},l,n}^{\ast}=\mathbf{v}_{\mathcal{S},l,n}^{\ast}(\mathbf{v}_{\mathcal{S},l,n}^{\ast})^H$ is an asymptotically optimal solution of Problem~\ref{BF1} at large $M$, where
\small{\begin{align}
    &\mathbf{v}_{\mathcal{S},l,n}^{\ast}\nonumber\\
    &=\frac{\sum\nolimits_{k\in\mathcal{K}_{\mathcal{S},l}}\frac{1}{\sqrt{\beta_k}}\mathbf{h}_{n,k}}{\left\|\sum\nolimits_{k\in\mathcal{K}_{\mathcal{S},l}}\frac{1}{\sqrt{\beta_k}}\mathbf{h}_{n,k}\right\|_2} \sqrt{\frac{M\sigma^2}{\min_{k\in\mathcal{K}_{\mathcal{S},l}}
\beta_k\frac{\left|\sum\nolimits_{j\in\mathcal{K}_{\mathcal{S},l}}\frac{\mathbf{h}_{n,k}^H\mathbf{h}_{n,j}}{\sqrt{\beta_j}}\right|^2}{\left\|\sum\nolimits_{j\in\mathcal{K}_{\mathcal{S},l}}\frac{\mathbf{h}_{n,j}}{\sqrt{\beta_j}}\right\|_2^2}}}.
\label{thm2-v}
\end{align}}
\end{Thm}


Substituting $Q_{\mathcal{S},l,n}^{\dagger(M)}=\text{trace}(\mathbf{V}_{\mathcal{S},l,n}^{\ast})$  into Problem~\ref{EP1}
and  \textcolor{black}{adopting} \textcolor{black}{the continuous relaxation and}  KKT conditions as in \cite{guo_ofdma},
we can obtain an asymptotically optimal solution of Problem~\ref{P1}, whose form is analogous to that in Lemma~1 in \cite{guo_ofdma}. Please refer to \cite{report} for details. \textcolor{black}{The asymptotically optimal solution  can achieve competitive performance \textcolor{black}{at large} $M$.}

\subsection{Suboptimal Solution in General Case}\label{sec:p1-dc}
In the general case (of arbitrary $M$),
\textcolor{black}{a low-complexity algorithm is developed} to obtain a suboptimal solution of Problem~\ref{P1} using  \textcolor{black}{continuous} relaxation and  DC programming.


\textcolor{black}{First, by \textcolor{black}{replacing} the constraints in \eqref{cst:mu1} of Problem~1 with  the following constraints:
\begin{equation}
\mu_{\mathcal{S},l,n}\geq 0,~~\mathcal{S}\in\mathcal{I},l\in\mathcal{L}_{\mathcal{S}},n\in\mathcal{N}, \label{const:ofdma-relax}
\end{equation}
\textcolor{black}{the relaxed version of Problem~\ref{P1}
involving only continuous optimization variables is obtained.}
Next, by the change of variables
$\mathbf{W}_{\mathcal{S},l,n}\triangleq\sqrt{\eta_{\mathcal{S},l,n} \mu_{\mathcal{S},l,n}}\mathbf{w}_{n}$,
the constraints in \eqref{cst:minrate}, \eqref{cst:eta} and \eqref{cst:w} can be equivalently transformed to the following \textcolor{black}{constraints}:}
\begin{align}
  &\mu_{\mathcal{S},l,n}\left(2^{\frac{c_{\mathcal{S},l,n} }{B\mu_{\mathcal{S},l,n} }}-1\right)-\frac{\beta_k|\mathbf{h}_{n,k}^H\mathbf{W}_{\mathcal{S},l,n}|^2}{M\sigma^2}\leq 0,\nonumber\\
  &~~~~~~~~~~~~~~~~~~~~~~~~~\mathcal{S}\in\mathcal{I},l\in\mathcal{L}_{\mathcal{S}},k\in\mathcal{K}_{\mathcal{S},l},n\in\mathcal{N}.\label{cst:minraten}
\end{align}
Therefore, \textcolor{black}{we can equivalently convert the relaxed continuous problem of Problem~\ref{P1} to}:
\begin{Prob}[DC Problem of Relaxed Problem~\ref{P1}]\label{RP1}
\begin{align}
\setlength{\abovedisplayskip}{0.2cm}
\setlength{\belowdisplayskip}{0.2cm}
 \min_{\mathbf{W},\bm{\mu},\mathbf{c} }~&\frac{1}{M}\sum\nolimits_{n\in\mathcal{N}}\sum\nolimits_{\mathcal{S}\in\mathcal{I}}\sum\nolimits_{l\in\mathcal{L}_{\mathcal{S}}}\|\mathbf{W}_{\mathcal{S},l,n} \|^2
 \nonumber\\
    \mathrm{s.t.} ~~
    &\eqref{cst:mu2},~\eqref{cst:c},~\eqref{cst:averate},~\eqref{const:ofdma-relax},~\eqref{cst:minraten}.\nonumber
\end{align}
\end{Prob}

Note that the objective function of Problem~\ref{RP1}
and the constraints in \eqref{cst:mu2},~\eqref{cst:c},~\eqref{cst:averate}, and \eqref{const:ofdma-relax} are all convex.
\textcolor{black}{Besides,
each constraint in \eqref{cst:minraten}}
can be regarded as a difference of two convex functions,
i.e., $\mu_{\mathcal{S},l,n}\left(2^{\frac{c_{\mathcal{S},l,n} }{B\mu_{\mathcal{S},l,n} }}-1\right)$ and $\frac{\beta_k|\mathbf{h}_{n,k}^H\mathbf{W}_{\mathcal{S},l,n}|^2}{M\sigma^2}$.
Thus, Problem~\ref{RP1} is a \textcolor{black}{standard} DC programming and can be handled by using the DC algorithm \textcolor{black}{\cite{dc}}. \textcolor{black}{In particular, we solve} a sequence of convex approximations of Problem~\ref{RP1} iteratively, each of which is obtained by \textcolor{black}{linearizing
 the concave function,
 i.e., $-\frac{\beta_k|\mathbf{h}_{n,k}^H\mathbf{W}_{\mathcal{S},l,n} |^2}{M\sigma^2}$} in \eqref{cst:minraten}.
\textcolor{black}{Specifically, at the $t$-th iteration,  the convex approximation of Problem~\ref{RP1}
is given below.
\begin{Prob}[Convex Approximation at
$t$-th Iteration]\label{CRP1}
\begin{align}
\setlength{\abovedisplayskip}{0.2cm}
\setlength{\belowdisplayskip}{0.2cm}
 &E^{(t)}\triangleq\min_{\mathbf{W},\bm{\mu},\mathbf{c} }~\frac{1}{M}\sum\nolimits_{n\in\mathcal{N}}\sum\nolimits_{\mathcal{S}\in\mathcal{I}}\sum\nolimits_{l\in\mathcal{L}_{\mathcal{S}}}\|\mathbf{W}_{\mathcal{S},l,n} \|^2 \nonumber\\
    &\mathrm{s.t.} ~
    \eqref{cst:mu2},~\eqref{cst:c},~\eqref{cst:averate},~\eqref{const:ofdma-relax},~\eqref{longrate},\nonumber
\end{align}
where \eqref{longrate} is shown at the top of the next page. Let $(\mathbf{W}^{(t)} ,\bm{\mu}^{(t)} ,\mathbf{c}^{(t)} )$ denote an optimal solution.
\end{Prob}}

\begin{figure*}[!t]
\vspace*{+0.2cm}
\small{\begin{align}
&\mu_{\mathcal{S},l,n}\left(2^{\frac{c_{\mathcal{S},l,n} }{B\mu_{\mathcal{S},l,n} }}-1\right)-\frac{2\beta_kR\left\{(\mathbf{W}^{(t-1)}_{\mathcal{S},l,n} )^H\mathbf{h}_{n,k} \mathbf{h}_{n,k}^H\mathbf{W}_{\mathcal{S},l,n}\right\}}{M\sigma^2}+\frac{|\mathbf{h}_{n,k}^H\mathbf{W}^{(t-1)}_{\mathcal{S},l,n} |^2}{M\sigma^2}\leq0,~\mathcal{S}\in\mathcal{I},l\in\mathcal{L}_{\mathcal{S}},k\in\mathcal{K}_{\mathcal{S},l},n\in\mathcal{N}.\label{longrate}
\end{align}
}
\vspace*{-0.5cm}
\end{figure*}


\textcolor{black}{Since Problem~\ref{CRP1} is a convex problem, we can solve it  using standard convex optimization techniques.}
According to \cite{dc}, \textcolor{black}{for any initial point which is a feasible solution of Problem~\ref{RP1}},
\textcolor{black}{as $t\rightarrow\infty$, $(\mathbf{W}^{(t)} ,\bm{\mu}^{(t)} ,\mathbf{c}^{(t)} )\rightarrow(\mathbf{W}^{(\infty)} ,\bm{\mu}^{(\infty)} ,\mathbf{c}^{(\infty)} )$, which
 is a stationary point of the relaxed Problem~\ref{P1}, and $E^{(t)}\rightarrow E^{(\infty)}$.
 \textcolor{black}{Note that $\bm{\mu}^{(\infty)}$ may not be binary, and hence
 $(\mathbf{W}^{(\infty)} ,\bm{\mu}^{(\infty)} ,\mathbf{c}^{(\infty)} )$ may not be a feasible solution of Problem~\ref{P1}.
 By the KKT conditions,  \textcolor{black}{an optimal solution of  Problem~\ref{CRP1}
 for the $t^{\diamond}$-the iteration can be obtained,}  where $t^{\diamond}$ satisfies} some
 convergence criteria. We shall show that the optimal solution provides binary subcarrier assignment under a mild condition, and \textcolor{black}{hence we can treat it} as a suboptimal solution of Problem~\ref{P1}.}

 Let $\bm{\lambda}_{\mathcal{S},l,n}\triangleq(\lambda_{\mathcal{S},l,n,k})_{k\in\mathcal{K}_{\mathcal{S},l}}$.
For all $\mathcal{S}\in\mathcal{I}$, $l\in\mathcal{L}_{\mathcal{S}}$ and
$n\in\mathcal{N}$, define:
\small{\begin{align}
&G_{\mathcal{S},l,n}(\gamma_{\mathcal{S},l},\bm{\lambda}_{\mathcal{S},l,n})\triangleq
\gamma_{\mathcal{S},l}\log_2\frac{\gamma_{\mathcal{S},l}}
{\ln2\sum\nolimits_{k\in\mathcal{S}}\lambda_{\mathcal{S},l,n,k}}-\frac{\gamma_{\mathcal{S},l}B}{\ln2}\nonumber\\
&~~~~~~~~~~~~~~~~~~~~~~~~+\sum\nolimits_{k\in\mathcal{S}}\lambda_{\mathcal{S},l,n,k},\label{cst:dc-g}\\
&\mu_{\mathcal{S},l,n}(\gamma_{\mathcal{S},l},\bm{\lambda}_{\mathcal{S},l,n})\nonumber\\
&=\begin{cases}
             1, ~ (\mathcal{S},l)=\argmax\limits_{\mathcal{S}'\in\mathcal{I},l'\in\mathcal{L}_{\mathcal{S}}}G_{\mathcal{S}',l',n}(\gamma_{\mathcal{S}',l'},\bm{\lambda}_{\mathcal{S}',l',n}) \\
             0, ~ \text{otherwise},
             \end{cases}\label{general-mu}\\
&c_{\mathcal{S},l,n}(\gamma_{\mathcal{S},l},\bm{\lambda}_{\mathcal{S},l,n})\nonumber\\
&=\mu_{\mathcal{S},l,n}(\gamma_{\mathcal{S},l},\bm{\lambda}_{\mathcal{S},l,n})B\left[\log_2\frac{\gamma_{\mathcal{S},l}}{\ln2\sum\limits_{k\in\mathcal{S}}\lambda_{\mathcal{S},l,n,k}}\right]^{+},\label{general-c}
\end{align}}
\noindent \noindent \normalsize{and $\mathbf{W}_{\mathcal{S},l,n}(\gamma_{\mathcal{S},l},\bm{\lambda}_{\mathcal{S},l,n})$ in \eqref{general_w} (as shown at the top of the next
page).}
\begin{figure*}[!t]
\small{\begin{align}
\mathbf{W}_{\mathcal{S},l,n}(\gamma_{\mathcal{S},l},\bm{\lambda}_{\mathcal{S},l,n})=\frac{\mu_{\mathcal{S},l,n}(\gamma_{\mathcal{S},l},\bm{\lambda}_{\mathcal{S},l,n})\sum\nolimits_{k\in\mathcal{S}}\lambda_{\mathcal{S},l,n,k}\beta_k(\mathbf{W}^{(\infty)}_{\mathcal{S},l,n} )^H\mathbf{h}_{n,k} \mathbf{h}_{n,k}^H
\sum\nolimits_{k\in\mathcal{S}}\lambda_{\mathcal{S},l,n,k}\beta_k|\mathbf{h}_{n,k}^H\mathbf{W}^{(\infty)}_{\mathcal{S},l,n} |^2}
{||\sum\nolimits_{k\in\mathcal{S}}\lambda_{\mathcal{S},l,n,k}\beta_k(\mathbf{W}^{(\infty)}_{\mathcal{S},l,n} )^H\mathbf{h}_{n,k} \mathbf{h}_{n,k}^H||^2},\label{general_w}
\end{align}
} 
\vspace*{-0.5cm}
\end{figure*}
Let $\gamma_{\mathcal{S},l}^{\infty}$ and $\bm{\lambda}_{\mathcal{S},l,n}^{\infty}$ denote the roots of \eqref{eq:lambda3}  (as shown at the top of the next page) and
$\sum\limits_{n\in\mathcal{N}}c_{\mathcal{S},l,n}(\gamma_{\mathcal{S},l},\bm{\lambda}_{\mathcal{S},l,n})=|\mathcal{P}_{\mathcal{S}}|D_l.$

\begin{figure*}[!t]
\small{\begin{align}
&\mu_{\mathcal{S},l,n}(\gamma_{\mathcal{S},l},\bm{\lambda}_{\mathcal{S},l,n})\left(2^{\frac{c_{\mathcal{S},l,n}(\gamma_{\mathcal{S},l},\bm{\lambda}_{\mathcal{S},l,n})}{B\mu_{\mathcal{S},l,n}(\gamma_{\mathcal{S},l},\bm{\lambda}_{\mathcal{S},l,n})}}-1\right)-\frac{2\beta_kR\left\{(\mathbf{W}^{(\infty)}_{\mathcal{S},l,n} )^H\mathbf{h}_{n,k} \mathbf{h}_{n,k}^H\mathbf{W}_{\mathcal{S},l,n}(\gamma_{\mathcal{S},l},\bm{\lambda}_{\mathcal{S},l,n})\right\}}{M\sigma^2}+\frac{|\mathbf{h}_{n,k}^H\mathbf{W}^{(\infty)}_{\mathcal{S},l,n} |^2}{M\sigma^2}=0,\label{eq:lambda3}
\end{align}
}\hrulefill
\end{figure*}

\begin{Cla}[Optimal Solution of Problem~\ref{CRP1}
 for $t^{\diamond}$]\label{lem:opt-ofdma2}
\textcolor{black}{Suppose that there exists a unique pair $(\mathcal{S}_n,l_n)$ such that $G_{\mathcal{S}_n,l_n,n}(\gamma_{\mathcal{S}_n,l_n}^{\diamond},\bm{\lambda}_{\mathcal{S}_n,l_n,n}^{\diamond})=\max\limits_{\mathcal{S}\in\mathcal{I},l\in\mathcal{L}_{\mathcal{S}}}
G_{\mathcal{S},l,n}(\gamma_{\mathcal{S},l}^{\diamond},\bm{\lambda}_{\mathcal{S},l,n}^{\diamond}),$  for all $n\in\mathcal{N}$.}
Then,
an optimal solution of Problem~\ref{CRP1}  for $t^{\diamond}$
 is given by
 $\mathbf{W}_{\mathcal{S},l,n}^{\diamond} =\mathbf{W}_{\mathcal{S},l,n}(\gamma_{\mathcal{S},l}^{\diamond},\bm{\lambda}_{\mathcal{S},l,n}^{\diamond})$,
 $\mu_{\mathcal{S},l,n}^{\diamond}  =\mu_{\mathcal{S},l,n}(\gamma_{\mathcal{S},l}^{\diamond},\bm{\lambda}_{\mathcal{S},l,n}^{\diamond})$
 and
 $c_{\mathcal{S},l,n}^{\diamond}  =c_{\mathcal{S},l,n}(\gamma_{\mathcal{S},l}^{\diamond},\bm{\lambda}_{\mathcal{S},l,n}^{\diamond})$.
\end{Cla}

Note that the optimal solution given in Claim~1  \textcolor{black}{guarantees} binary subcarrier assignments.
As illustrated in
\cite{guo_ofdma}, the condition in Claim~1 can be easily satisfied \textcolor{black}{in practical systems}. Note that $\gamma_{\mathcal{S},l}^{\diamond}$ and $\bm{\lambda}_{\mathcal{S},l,n}^{\diamond}$ can be obtained using a subgradient method,
and an optimal solution of the convex approximation problem of Problem~\ref{RP1}
can be obtained.
The details for obtaining a suboptimal solution $(\bm{\mu}^{\diamond},\bm{\eta}^{\diamond},\mathbf{c}^{\diamond},\mathbf{w}^{\diamond})$
of Problem~1 are summarized in Algorithm~1.
\vspace*{-0.2cm}
\begin{algorithm}
    \caption{\small{Suboptimal Solution of Problem~\ref{P1} for General Case}}
\begin{footnotesize}
     \begin{algorithmic}[1]
     \STATE Find a random feasible point of Problem~\ref{RP1} as the initial point $(\mathbf{W}^{(0)},\bm{\mu}^{(0)},\mathbf{c}^{(0)})$, and set $t=0$;
     \REPEAT
     \STATE Set $t=t+1$;
     \STATE Obtain $(\mathbf{W}^{(t)},\bm{\mu}^{(t)},\mathbf{c}^{(t)})$
     by  solving Problem~\ref{CRP1} using standard convex optimization techniques;
     \UNTIL{convergence criteria are met}
         \STATE Set $t^{\diamond}=t$, initialize $\bm{\gamma}^{(1)}$ and $\bm{\lambda}^{(1)}$, and set $i=0$;
     \REPEAT
     \STATE Set $i=i+1$;
           \STATE For all $\mathcal{S}\in\mathcal{I},l\in\mathcal{L}_{\mathcal{S}}$ and $n\in\mathcal{N}$, compute $G_{\mathcal{S},l,n}(\gamma_{\mathcal{S},l}^{(i)},\bm{\lambda}_{\mathcal{S},l,n}^{(i)})$, $\mu_{\mathcal{S},l,n}(\gamma_{\mathcal{S},l}^{(i)},\bm{\lambda}_{\mathcal{S},l,n}^{(i)})$,
           $c_{\mathcal{S},l,n}(\gamma_{\mathcal{S},l}^{(i)},\bm{\lambda}_{\mathcal{S},l,n}^{(i)})$  and $\mathbf{W}_{\mathcal{S},l,n}(\gamma_{\mathcal{S},l}^{(i)},\bm{\lambda}_{\mathcal{S},l,n}^{(i)})$  according to \eqref{cst:dc-g}, \eqref{general-mu}, \eqref{general-c} and
           \eqref{general_w}, respectively;
           \STATE For all $\mathcal{S}\in\mathcal{I}$, $l\in\mathcal{L}_{\mathcal{S}}$, $n\in\mathcal{N}$ and $k\in\mathcal{K}_{\mathcal{S},l}$, compute $\lambda^{(i+1)}_{\mathcal{S},l,n,k}$ according to \eqref{alg-t1} (as shown at the top of the next page),
where $\delta^{(i)}>0$, $i=1,2,\ldots$ satisfy
\begin{equation}
\setlength{\abovedisplayskip}{0.2cm}
\setlength{\belowdisplayskip}{0.2cm}
 \sum\nolimits_{i=0}^{\infty}(\delta^{(i)})^2<\infty,~\sum\nolimits_{i=0}^{\infty}\delta^{(i)}=\infty,~\text{lim}_{i\rightarrow\infty}\delta^{(i)}=0;\label{cst:step3}
 \end{equation}
\STATE For all $\mathcal{S}\in\mathcal{I}$ and $l\in\mathcal{L}_{\mathcal{S}}$, compute $\gamma^{(i+1)}_{\mathcal{S},l}$ according to
\begin{align}
\setlength{\abovedisplayskip}{0.2cm}
\setlength{\belowdisplayskip}{0.2cm}
&\gamma_{\mathcal{S},l}^{(i+1)}\nonumber\\
&=\left[\gamma_{\mathcal{S},l}^{(i)}-\delta^{(i)}
\left(\sum\nolimits_{n\in\mathcal{N}}c_{\mathcal{S},l,n}(\gamma_{\mathcal{S},l}^{(i)},\bm{\lambda}_{\mathcal{S},l,n}^{(i)})-|\mathcal{P}_{\mathcal{S}}|D_l\right)\right]^+,\nonumber
\end{align}
\textcolor{black}{where $\delta^{(i)}>0$, $i=1,2,\ldots$
 satisfy \eqref{cst:step3};}
     \UNTIL{convergence criteria are met}
     \STATE Set $\bm{\gamma}^{\diamond}(\mathbf{h})=\bm{\gamma}^{(i)}$ and $\bm{\lambda}^{\diamond}(\mathbf{h})=\bm{\lambda}^{(i)}$;
     \STATE For all $\mathcal{S}\in\mathcal{I},~l\in\mathcal{L}_{\mathcal{S}}$ and $n\in\mathcal{N}$, set
     $\mu_{\mathcal{S},l,n}^{\diamond}=\mu_{\mathcal{S},l,n}(\gamma_{\mathcal{S},l}^{\diamond},\bm{\lambda}_{\mathcal{S},l,n}^{\diamond})$,
      $\eta_{\mathcal{S},l,n}^{\diamond}=\|\mathbf{W}_{\mathcal{S},l,n}(\gamma_{\mathcal{S},l}^{\diamond},\bm{\lambda}_{\mathcal{S},l,n}^{\diamond})\|_2$, $c_{\mathcal{S},l,n}^{\diamond}=c_{\mathcal{S},l,n}(\gamma_{\mathcal{S},l}^{\diamond},\bm{\lambda}_{\mathcal{S},l,n}^{\diamond})$
      and $\mathbf{w}_{n}^{\diamond}=\sum\nolimits_{\mathcal{S}\in\mathcal{I}}\sum\nolimits_{l\in\mathcal{L}_{\mathcal{S}}}\mu_{\mathcal{S},l,n}(\gamma_{\mathcal{S},l}^{\diamond},\bm{\lambda}_{\mathcal{S},l,n}^{\diamond})\frac{\mathbf{W}_{\mathcal{S},l,n}(\gamma_{\mathcal{S},l}^{\diamond},\bm{\lambda}_{\mathcal{S},l,n}^{\diamond})}{\eta_{\mathcal{S},l,n}^{\diamond}}$.
    \end{algorithmic}
    \end{footnotesize}\label{alg:opt-SP1}
\end{algorithm}

\begin{figure*}[!t]
\vspace*{-0.1cm}
\small{\begin{align}
&\lambda_{\mathcal{S},l,n,k}^{(i+1)}=\left[\lambda_{\mathcal{S},l,n,k}^{(i)}-\delta^{(i)}
\left(\mu_{\mathcal{S},l,n}(\gamma^{(i)}_{\mathcal{S},l},\bm{\lambda}^{(i)}_{\mathcal{S},l,n})\left(2^{\frac{c_{\mathcal{S},l,n}(\gamma^{(i)}_{\mathcal{S},l},\bm{\lambda}^{(i)}_{\mathcal{S},l,n})}{B\mu_{\mathcal{S},l,n}(\gamma^{(i)}_{\mathcal{S},l},\bm{\lambda}^{(i)}_{\mathcal{S},l,n})}}-1\right)
-\frac{2\beta_kR\left\{(\mathbf{W}^{(t^{\diamond})}_{\mathcal{S},l,n})^H\mathbf{h}_{n,k}(\mathbf{h}_{n,k})^H\mathbf{W}_{\mathcal{S},l,n}(\gamma^{(i)}_{\mathcal{S},l},\bm{\lambda}^{(i)}_{\mathcal{S},l,n})\right\}}{M\sigma^2}\right.\right.\nonumber\\
&\left.\left.+\frac{|(\mathbf{h}_{n,k})^H\mathbf{W}^{(t^{\diamond})}_{\mathcal{S},l,n}|^2}{M\sigma^2}\right)\right]^+,
\label{alg-t1}
\end{align}
} \hrulefill
\end{figure*}

\vspace*{-0.05cm}
\section{Numerical Results}
\vspace*{-0.14cm}
In this section, we compare the proposed solutions
with \textcolor{black}{two} baseline schemes.
Baseline~1 \textcolor{black}{serves} $K$
 users separately (i.e., adopts unicast), and adopts the normalized maximum ratio transmission (MRT) beamformer for each user on each subcarrier.
Baseline~2 jointly considers the FoVs of all users \textcolor{black}{(i.e., adopts multicast for a message, if there exists a multicast opportunity)} as in this paper,
and adopts the normalized MRT beamformer for a massage on each subcarrier  obtained based on
 the channel  matrix of all users requiring this message on each subcarrier \cite{8392787}.
Then, for each baseline scheme,
the optimal subcarrier, power
and rate allocation is obtained by solving Problem~\ref{EP1}
 for the respective MRT using the method proposed in \cite{guo_ofdma}.
In this simulation, we set $\beta_k=1$ for all $k\in\mathcal{K}$, $F_h=F_v=100^{\circ}$, $U_h\times U_v=30\times15$, $B =39$ kHz, $N=64$,  $\sigma^2 = 10^{-9}$ W, and assume
$\mathbf{h}_{n,k}$, \textcolor{black}{$n\in\mathcal{N}$, $k\in\mathcal{K}$} are randomly and independently distributed according to $\mathcal{CN}(0,\mathbf{1}_{M\times M})$.
We consider the 360 VR video sequence $Venice$  \cite{vr-sequence1}.
The 360 VR video encoder named Kvazaar is adopted. Set $L=5$, and choose $D_l,l\in\mathcal{L}$ as in \cite{long}.
Given the viewing direction of a user, the associated FoV of size  $F_h\times F_v$ can be determined.
\textcolor{black}{To avoid delay in view switching,}
extra $15^{\circ}$ in the four directions of the requested FoV \textcolor{black}{is transmitted}, determining $\mathcal{G}_k$  \textcolor{black}{for each user $k\in\mathcal{K}$} \cite{guo_ofdma,guo_tdma}.
For any $\mathcal{G}_k,k\in\mathcal{K}$, we evaluate the average power
 over 100 random \textcolor{black}{realizations of} system channel states.

\begin{figure}[t]
\vspace*{-0.3cm}
\begin{center}
 \subfigure[\small{Average power versus $K$. $M=4$, $\mathbf{r}=(2,2,3,3,4)$.}]
 {\resizebox{4.15cm}{!}{\includegraphics{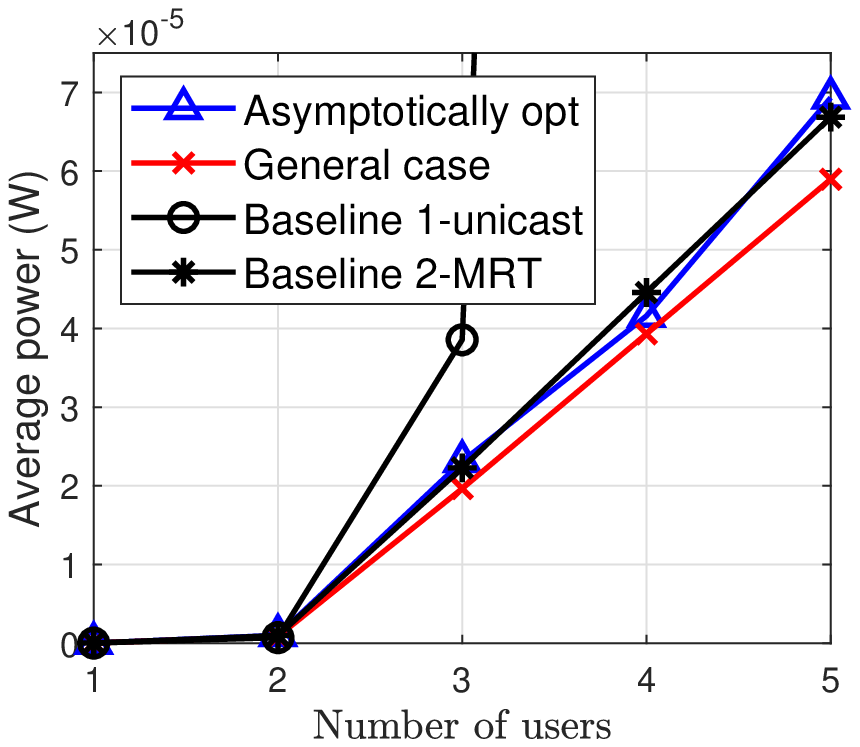}}}
  \subfigure[\small{Average power  versus $M$. $K=4$, $\mathbf{r}=(2,3,3,4)$.}]
 {\resizebox{4.15cm}{!}{\includegraphics{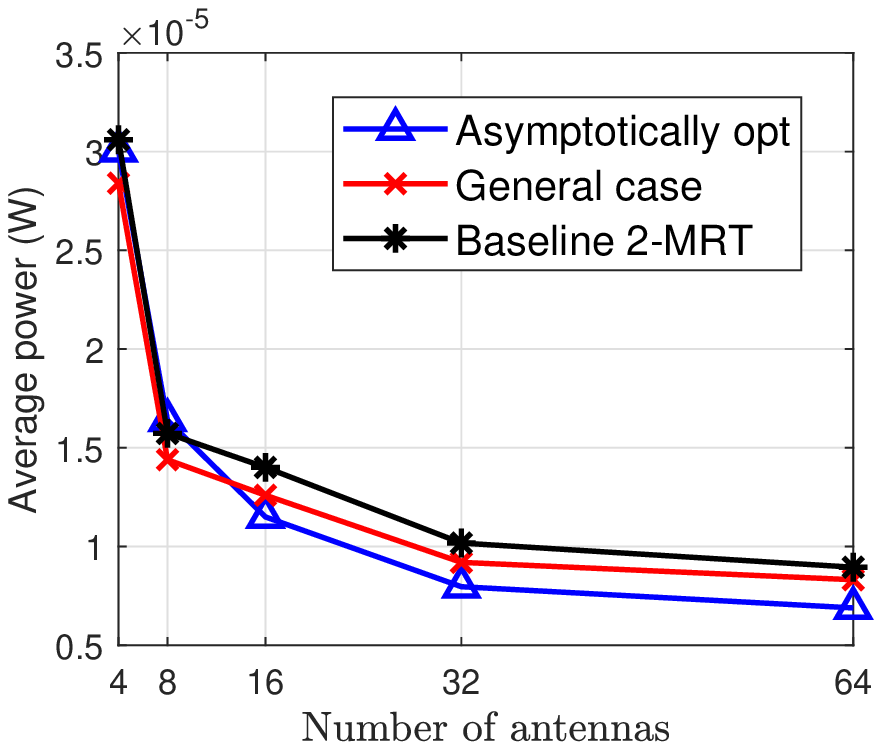}}}
 \end{center}
 \vspace*{-0.25cm}
   \caption{\small{Average power versus $K$ and $M$. }}
   \label{fig:zipf}
\vspace*{-0.35cm}
\end{figure}

\begin{figure}[t]
\begin{center}
 \includegraphics[width=2.5in]{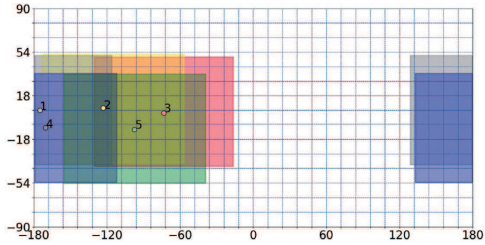}
\vspace*{-0.07cm}
  \end{center}
   \caption{\small{Viewing directions and corresponding FoVs of $5$ users \cite{vr-sequence1}. }}
   \label{fig:fov}
\vspace*{-0.30cm}
\end{figure}

First, we evaluate the average power over 1,000 random choices for
the viewing directions of 1-5 users from 30 users in \cite{vr-sequence1}.
Fig.~\ref{fig:zipf}~(a) illustrates the average power versus the number of users $K$.
We can see that the  average powers of the proposed solutions and \textcolor{black}{baseline schemes} increase with $K$, as the transmission load increases with $K$.
\textcolor{black}{
Given the unsatisfactory performance of Baseline~1,
 we no longer compare with it in the remaining figures.}
Fig.~\ref{fig:zipf}~(b)  illustrates the average power versus the number of antennas $M$.
We can see  the  powers achieved by the proposed solutions and baseline
schemes
decrease with $M$. \textcolor{black}{Besides, by Fig.~\ref{fig:zipf}~(b), we can observe} that the proposed asymptotically optimal solution achieves better performance than Baseline~2 when the number of antennas is larger than 8.
\begin{figure}[t]
\begin{center}
 {\resizebox{4.15cm}{!}{\includegraphics{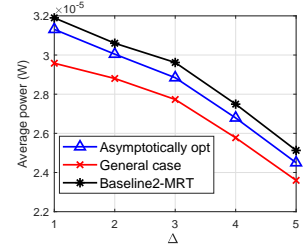}}}
 \end{center}
 \vspace*{-0.3cm}
   \caption{\small{Average power versus $\Delta$.  $K=5$, $M=4$, $\mathbf{r}=(2,2,3,3,4)$. }}
   \label{fig:fovc}
\vspace*{-0.50cm}
\end{figure}
Next, we show the impact of concentration
 of the viewing directions of  all users.
 We choose the viewing
 directions of 5 users out of 30 users in
 \cite{vr-sequence1}, i.e., $(\nu_k,\gamma_k)_{k\in\{1,\ldots,5\}}$,
 \textcolor{black}{ as shown in Fig.~\ref{fig:fov}.
 Based on the chosen viewing directions,
 we consider five sets of viewing directions, i.e.,
 $(\nu_1+\Delta,\gamma_1),(\nu_2+\Delta,\gamma_2),(\nu_3,\gamma_3),(\nu_4-\Delta,\gamma_4),$ and $(\nu_5-\Delta,\gamma_5)$,  $\Delta=1,\ldots,5$,
 and evaluate the corresponding average powers.
Note that $\Delta$ reflects the concentration of the viewing directions of the 5 users.
In particular, the concentration increases with $\Delta$.
Fig.~\ref{fig:fovc} shows the average power versus the concentration parameter $\Delta$. \textcolor{black}{It can be observed} that  the average
power of each multicast scheme  decreases with $\Delta$, since multicast
opportunities increase with $\Delta$.}
Finally, from Fig.~\ref{fig:zipf} and Fig.~\ref{fig:fovc}, \textcolor{black}{it can be observed} that the
proposed solutions \textcolor{black}{perform better than} the  baseline schemes. Specifically,
\textcolor{black}{the proposed solutions outperform  Baseline~1,
as they achieve
a  higher spectral efficiency utilizing multicast opportunities.
The proposed solutions outperform  Baseline~2, as they
carefully choose beamforming vectors.}

\vspace{-0.05cm}
\section{Conclusion}
\vspace{-0.14cm}
In this paper, we studied optimal transmission of a multi-quality tiled 360 VR video to multiple users in an MIMO-OFDMA system.
\textcolor{black}{We minimized the total transmission power by optimizing} the beamforming vector and subcarrier, transmission power and rate allocation.
\textcolor{black}{This is a  challenging mixed
discrete-continuous optimization problem.
We} obtained an asymptotically optimal solution in the case of a large antenna array, and a suboptimal solution in the general case.
Finally, numerical results showed that the proposed solutions achieve significant gains over
existing schemes.
%


\end{document}